\documentclass[11pt, oneside]{article}   	
\usepackage[margin=3cm]{geometry}                		
\usepackage{amsmath,amssymb}          
\usepackage{graphicx}
\usepackage{float}
\usepackage{psfrag,epsf}
\usepackage{pgf,tikz}
\usepackage{epstopdf}	
\usepackage[T1]{fontenc}
\usepackage[toc,page,header]{appendix}
\usepackage{minitoc}
\usepackage{subfiles}
\usepackage{chapterbib}
\RequirePackage{amsthm,amsmath}
\usepackage{natbib}
\usepackage{url}            
\usepackage{booktabs}       
\usepackage{amsfonts}       
\usepackage{nicefrac}       
\usepackage{microtype}      
\usepackage{multirow}
\usepackage{threeparttable}
\usepackage{siunitx}
\usepackage{caption}
\usepackage{subcaption}
\usepackage{ragged2e}
\usepackage{hyperref}
\usepackage{authblk}
\hypersetup{colorlinks,linkcolor={blue},citecolor={blue},urlcolor={blue}}

\definecolor{myblue}{rgb}{0,0,0.75}

\usepackage{lineno}
\makeatletter
\def\makeLineNumberLeft{%
  \linenumberfont\llap{\hb@xt@\linenumberwidth{\LineNumber\hss}\hskip\linenumbersep}
  \hskip\columnwidth
  \rlap{\hskip\linenumbersep\hb@xt@\linenumberwidth{\hss\LineNumber}}\hss}
\leftlinenumbers
\makeatother

\newcommand{\bs}[1]{{\boldsymbol{#1}}}
\newcommand{\con}{{\,|\,}}

\newcommand{\bbP}{\mathbb{P}}
\newcommand{\bbE}{\mathbb{E}}
\newcommand{\bbR}{\mathbb{R}}
\newcommand{\T}{\textsc{T}}
\newcommand{\abs}[1]{\left\lvert#1\right\rvert}
\newcommand{\norm}[1]{\left\lVert#1\right\rVert}
\newcommand{\normal}{\text{N}}
\newcommand{\bernoulli}{\text{Bernoulli}}
\newcommand{\uniform}{\text{Uniform}}
\newcommand{\gammaprior}{\text{Gamma}}

\title{Hurdle Network Model With Latent Dynamic Shrinkage For Enhanced Edge Prediction in Zero-Inflated Directed Network Time Series}
\author[1]{Sandipan Pramanik\thanks{Corresponding Author. Email: \href{mailto:spraman4@jhu.edu}{spraman4@jhu.edu}}}
\author[2]{Raymond Robertson\thanks{Email: \href{mailto:robertson@tamu.edu}{robertson@tamu.edu}}}
\author[3]{Yang Ni\thanks{Email: \href{mailto:yni@stat.tamu.edu}{yni@stat.tamu.edu}}}
\affil[1]{Department of Biostatistics, Johns Hopkins University}
\affil[2]{The Bush School of Government and Public Service, Texas A\&M University}
\affil[3]{Department of Statistics, Texas A\&M University}

\date{\ }							

\begin{document}

\maketitle

\begin{abstract}
This article aims to model international trade relationships among 29 countries within the apparel industry from 1994 to 2013. Bilateral trade relationships can be described by a network, where nodes correspond to countries and directed edges indicate trade relationships (that is, whether the source node exported to the target node in a given year). Additionally, both node-specific covariates (e.g., GDP) and edge-specific covariates (e.g., labor provision) are available. This study focuses on two key aspects. First, there are at least three types of dependence: temporal dependence, network dependence, and dependence on covariates. Second, we identify the \textit{potential trade volume} as an important edge-specific covariate, which is intuitively useful in predicting trade relationships. However, potential trade volume is only observed for country pairs engaged in trade, making it challenging to account for.

We introduce the \textit{dynamic hurdle network (Hurdle-Net)} model for zero-inflated directed network time series to incorporate these key features effectively. It includes several novel components. First, by representing the zero-inflated time series as a paired binary and continuous network time series of edge occurrence and edge weights, it presents a hurdle model for networks that naturally accommodates zero inflation (sparse trade relationships) in edge weights. Second, it incorporates node-specific latent variables to capture network dependencies and their evolution over time. We employ the \textit{dynamic shrinkage process prior} on latent variables for this purpose. Third, recognizing that both network time series involve the same nodes, Hurdle-Net leverages a common latent mechanism underlying both binary and continuous network time series. Finally, it utilizes the generalized logistic function as a link function to establish a correspondence between the two network time series. This enables a parsimonious functional relationship between the edge occurrence and edge weight networks, and jointly models them in a hierarchical Bayesian framework. Compared to the \textit{independent} and \textit{static} modeling approaches, Hurdle-Net achieves superior model selection and improved estimation/prediction for international trade relationships. 
Simulation studies and the application to bilateral trade flow data illustrate its effectiveness and advantages.
\end{abstract}


\section{Introduction}\label{sec: Introduction}



Recent technological advancements in diverse scientific areas have significantly simplified the process of collecting data on numerous variable and their interactions over time. Some examples include functional connectivity networks among brain regions \citep{Tomasi2011}, 
interactions within social networks \citep{Wilson2009}, email communication networks \citep{diesner2005exploration}, citation networks among research articles or authors \citep{Yan2012}, and networks of co-purchased products \citep{Kafkas2021}. These data sets are instances of network time series, where each node represents an individual or actor in the network. An edge between two nodes indicates an interaction between the individuals or actors. In a network time series, the dependence within the network evolves and interactions between nodes vary over time.

This article is motivated by analyzing and predicting bilateral trade relationships among 29 countries in the apparel industry between 1994 and 2013. Over these 20 years, the data recorded trade occurrences between each country pair and the trade volume in the presence of a trade occurrence. The data can be represented as a paired time series: a binary directed network time series of trade occurrences, and a continuous directed network time series of nonzero trade volumes, which is zero-inflated as $30\%$ of the country pairs did not engage in trade. The data also recorded additional covariate information on countries (e.g., GDP and population) and country pairs (e.g., labor provision and distance), which are node-specific and edge-specific attributes, respectively.

Statistical modeling of networks has been an area of active research for many years, as highlighted by \cite{kim2018} in a comprehensive review. The methodologies in this field typically fall into two main categories based on whether they use latent variables to model network dependence or not. Stochastic block models and latent space models (LSMs) are two widely used frameworks utilizing latent variables. 
Stochastic block models assume that the nodes can be partitioned into different groups, with interaction probabilities between nodes depending on their group memberships \citep{HOLLAND1983}. It encourages more interactions within groups than between them, with the number of groups and individual memberships often being unknown. A key objective is estimating group memberships and interaction probabilities. LSMs assume node-specific latent attributes \citep{hoff2002}. Geometrically, the nodes are assumed to lie in an Euclidean space where the presence of an edge between two nodes depends on the similarity of their latent positions. This provides a geometric representation and interpretation of the network. Models that do not use latent variables, such as exponential random graph models \citep{ROBINS2007}, the quadratic assignment procedure \citep{KRACKHARDT1988}, and stochastic actor-oriented models \citep{Snijders1996}, directly assume a probability distribution on graphs without incorporating latent variables. 

A substantial body of literature has explored LSMs in many real-world applications \citep{hoff2002, hoff2004, ward2007, fletcher2011, krafft2012, ward2013, henry2019}. 
LSMs have been extended to model network time series data such as those based on the Gaussian random walk on latent positions \citep{sarkar2006, sarkar2007}. While these methods lay the foundation for modeling network time series, they can be improved in several aspects. For example, they often limit the time dependence to a Markov structure. Also, they are not well-suited for modeling zero-inflated weighted network data. To this, \cite{ward2013} considered an independent modeling approach where edge occurrences (binary network) and edge weights (continuous network) are modeled independently and sequentially at each time point. This relates to separate binary and continuous network regressions, implicitly assuming that the mechanisms for these two networks are independent. This substantially increases the number of parameters and can be inefficient when edge occurrence is unbalanced \citep{Japkowicz2002}. Moreover, they do not utilize information from both networks simultaneously, despite them relating to the same set of nodes.

We present the \textit{dynamic hurdle network (Hurdle-Net)} model to enhance edge occurrence prediction in zero-inflated directed network time series. The framework represents the data as paired binary and continuous directed network time series and incorporates several key innovations. First, we develop a hurdle model for network data that naturally accommodates zero inflation in edge weights. Second, we introduce node-specific latent variables to capture underlying network dependence among the nodes and model their temporal evolution. Third, recognizing that both network time series share the same nodes, Hurdle-Net posits a common latent mechanism underlying both networks to account for the shared relationship between edge occurrences and edge weights. Lastly, we employ the generalized logistic function as a general link between the two network time series, establishing a parsimonious and efficient joint modeling framework.


The network data at each time point can be represented as an asymmetric square matrix. The proposed latent formulation ensures that the inference from Hurdle-Net is invariant to the transpose/orientation of the data. We assume a dynamic shrinkage process (DSP) prior on latent positions to model their temporal evolution \citep{kowal2019}. 
DSP is a global-local continuous shrinkage prior, which allows a wide range of latent dynamic processes. 
Hurdle-Net also naturally incorporates covariates specific to individual nodes and node pairs. Compared to methods that do not incorporate trade-volume data (\textit{independent} model) or time-dependence (\textit{static} model), Hurdle-Net demonstrates enhanced model selection and improved estimation/prediction performance for edge occurrence, while maintaining similar or better performance for edge weight estimation/prediction. We demonstrate the effectiveness and superiority of the proposed method through both simulation studies and its application to bilateral trade flow data. Code for implementation


The rest of the article is organized as follows. Section~\ref{sec: Motivating Dataset} describes the motivating dataset. In Section~\ref{sec: Method}, we propose our methodology. Section~\ref{sec: Simulation Study} investigates the empirical properties of our method through simulation studies that mimic real-world applications. In Section~\ref{sec: Bilateral Trade Flows In The Apparel Industry}, we revisit the motivating dataset and apply the proposed model to analyze bilateral trade flows. Finally, Section~\ref{sec: Discussion} concludes with a discussion of our findings and future research opportunities.

\section{Motivating Dataset}\label{sec: Motivating Dataset}

\begin{figure}[!t]
     \centering
     \begin{subfigure}[b]{0.49\linewidth}
         \centering
         \includegraphics[width=.9\linewidth]{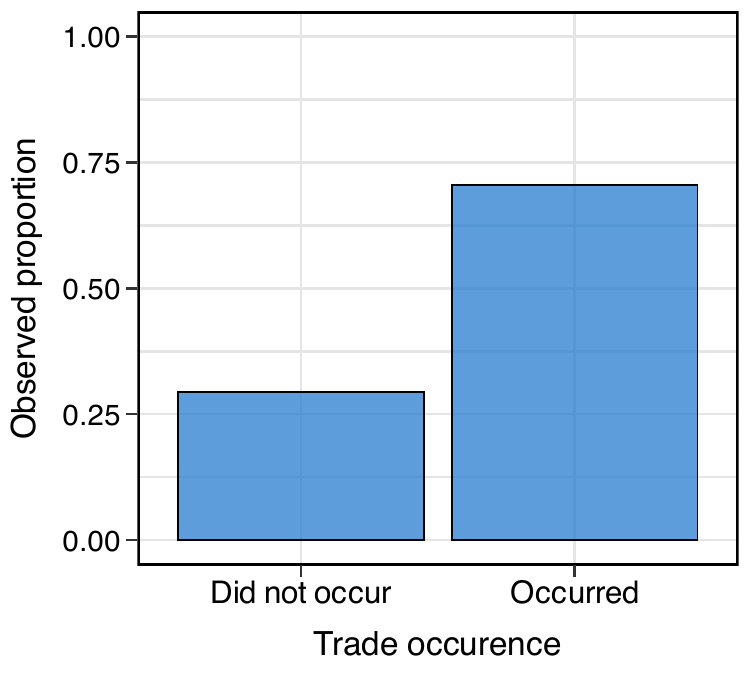}
         \caption{}
         \label{subfig: occurrence}
     \end{subfigure}
     \hfill
     \begin{subfigure}[b]{0.49\linewidth}
         \centering
         \includegraphics[width=.9\linewidth]{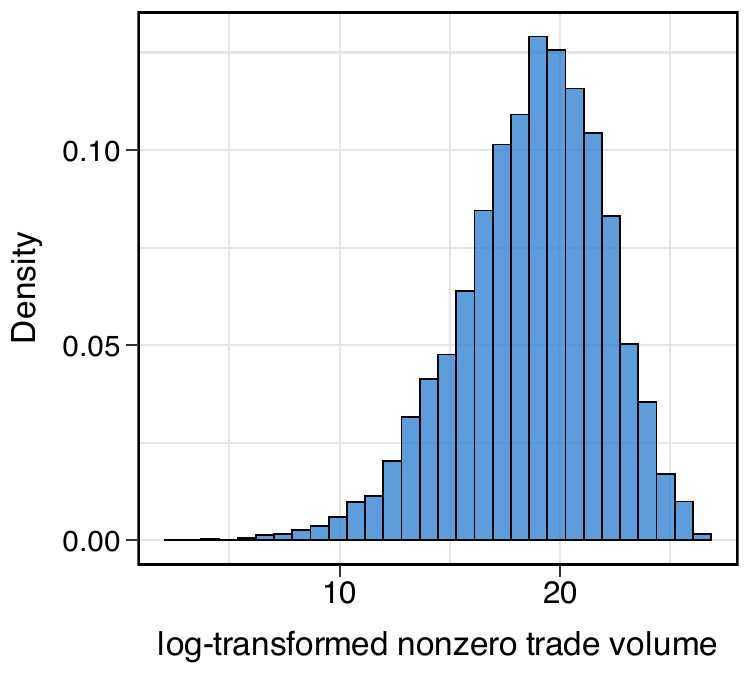}
         \caption{}
         \label{subfig: histogram nonzero}
     \end{subfigure}
   \caption{Overview of observed trade occurrences and nonzero trade volumes in the apparel industry from 1994 to 2013 among 29 countries. Panel~(a) shows the proportion of trade occurrence. Panel~(b) depicts the distribution of nonzero trade volumes after $\log$ transformation.}
   \label{fig: trade histogram}
\end{figure}

 The motivating dataset recorded bilateral trade flows among 29
countries in the apparel industry between 1994 and 2013. Figure~\ref{fig: trade histogram} offers an overview of the observed trade occurrences and nonzero trade volumes for trading country pairs. Approximately $30\%$ of country pairs do not have trades, and, for trading countries, the distribution of their trade volumes after the $\log$ transformation demonstrates an unimodal pattern. These characteristics collectively suggest a hurdle model mechanism that jointly accommodates the probability of attaining zero and the distribution of non-zero values. The dataset also recorded continuous (GDP, populations, areas, and distances between capitals) and binary covariates (regional trade agreements and labor provisions between two countries). Among them, distance, regional trade agreement, and labor provision are specific to each country pair. The rest of them are country-specific.


\section{Method}\label{sec: Method}

In this section, we introduce a dynamic hurdle network model for jointly modeling the temporal dynamics in both binary and continuous network time series. Although the proposed model is motivated by a specific application to bilateral trade flows, it is broadly applicable to zero-inflated network time series data. We begin by introducing the necessary notations, then illustrate the dynamic hurdle network model, discuss the role of the dynamic shrinkage process prior in modeling the latent evolution of network dependence, and provide the necessary specifics for Markov chain Monte Carlo (MCMC) implementation to conduct Bayesian inference.

\subsection{Notations}\label{subsec: Notations}

Suppose there are $n$ nodes in a directed graph. For any pair of distinct nodes $(i,j)$, let $\delta_{ijt}$ denote the presence $(\delta_{ijt}=1)$ or absence $(\delta_{ijt}=0)$ of an edge from node $i$ to $j$ at time $t$ (edge occurrence). Similarly, let $y_{ijt}$ represent a continuous measurement from node $i$ to $j$ at time $t$ (edge weight), which takes the value of zero if and only if $\delta_{ijt}=0$. Together, $\{\left( \bs{\Delta}_t , \mathbf{Y}_t \right)\}_{t=1}^T$ denotes the zero-inflated directed network time series where $\bs{\Delta}_t = (\delta_{ijt})$ is the $n \times n$ binary asymmetric adjacency matrix and $\mathbf{Y}_t = (y_{ijt})$ is the $n \times n$ continuous asymmetric adjacency matrix. Additionally, $\bs{x}_{it} \in \bbR^{p_1}$ denote $p_1$ covariates specific to node $i$, and $\bs{x}_{i \bullet j,t} \in \bbR^{p_2}$ denote $p_2$ covariates specific to the node pair $(i,j)$ (in our application, $\bs{x}_{i \bullet j,t} = \bs{x}_{j \bullet i,t}$). To simplify the notation, we define $\bs{x}_{ijt} = \left(\bs{x}_{it}, \bs{x}_{jt}, \bs{x}_{i \bullet j,t} \right)^\T$ as the vector of $p = 2 p_1 + p_2$ covariates associated with the interaction from node $i$ to node $j$ at time $t$.

\subsection{Dynamic Hurdle Network Model}\label{subsec: model}

Following \cite{hoff2002}, we adopt a conditional independence approach at each time point to model network dependence among nodes. Let $\mathbf{Z}_t = \left(\bs{z}_{1t},\ldots,\bs{z}_{nt}\right)^\T$, where $\bs{z}_{it} \in \bbR^{K}$ denotes the latent position of node $i$ at time $t$ in a $K$-dimensional Euclidean space, representing its unmeasured characteristics. Given $\mathbf{Z}_t$, the model assumes that the edge occurrence in the binary network from node $i$ to $j$ is independent of all other edge occurrences. Likewise, in the continuous network, the edge weight from $i$ to $j$ is assumed to be independent of all other edge weights when conditioned on $\mathbf{Z}_t$. Combined together, $\left(\delta_{ijt}, y_{ijt}\right)$ and $(\delta_{klt}, y_{klt})$ are conditionally independent given $\mathbf{Z}_t$ for any $(i,j) \neq (k,l)$.
    
Moreover, the edge occurrence probability from node $i$ to $j$ at time $t$ in the binary network is assumed to increase with the similarity of the latent variables $\bs{z}_{it}$ and $\bs{z}_{jt}$. Likewise, in the continuous network, the expected edge weight from node $i$ to $j$ at time $t$ given the presence of an edge is also assumed to increase with the similarity of the corresponding latent variables. 

Specifically, for all $t$ and $i \neq j$, we assume the following probabilistic models for both edge occurrences and edge weights given latent variables $\mathbf{Z}_t$:
\begin{align}
    & y_{ijt} \con \delta_{ijt} = 1   \overset{ind}{\sim} \normal \left( \bs{x}_{ijt}^\T \, \bs{\beta}_\text{C} + \text{L}_{ijt}, \sigma^2 \right),\label{eq: cont model}\\
    & \delta_{ijt}    \overset{ind}{\sim}\, \bernoulli \left( g \left(\bs{x}_{ijt}^\T \, \bs{\beta}_\text{P} + \text{L}_{ijt} \right) \right), \label{eq: probit model}
\end{align}
where $\normal(w,\sigma^2)$ denote the normal distribution with mean $w$ and variance $\sigma^2$, and $\bernoulli (q)$ denote the Bernoulli distribution with success probability $q$.
The parameter $\text{L}_{ijt}$ measures the similarity between $\bs{z}_{it}$ and $\bs{z}_{jt}$.
Parameters $\bs{\beta}_\text{C}$ and $\bs{\beta}_\text{P}$ quantify the effects of different covariates on the expected edge weight and edge occurrence probability. 
We refer to (\ref{eq: probit model}) as the \textit{binary model} and it models the edge occurrences. Conditioned on $\delta_{ijt}$, (\ref{eq: cont model}) models edge weights between node pairs for which edges are present. We refer to the latter as the \textit{continuous model}. Together (\ref{eq: cont model})--(\ref{eq: probit model}) specify a joint \textit{dynamic hurdle model} for zero-inflated directed network time series (Hurdle-Net hereafter). 


For the similarity $\text{L}_{ijt}$ between $\bs{z}_{it}$ and $\bs{z}_{jt}$, we propose the following formulation,

\begin{equation}\label{eq: latent term}
    \text{L}_{ijt} = \alpha \, \frac{\bs{z}_{it}^\T \bs{z}_{jt}}{\norm{\bs{z}_{jt}}} + (1-\alpha) \, \frac{\bs{z}_{it}^\T \bs{z}_{jt}}{\norm{\bs{z}_{it}}},
\end{equation}
where $\alpha \in [0,1]$ and $\norm{\cdot}$ is the Euclidean norm. Note that, the first term $\bs{z}_{it}^\T \bs{z}_{jt}/\norm{\bs{z}_{jt}} = \bs{a}_i \bs{v}_{it}^\T \bs{v}_{jt}$ where $\bs{v}_{it} = \bs{z}_{it}/\norm{\bs{z}_{it}}$ is used to quantify the similarity $\bs{v}_{it}^\T\bs{v}_{jt}$ between nodes $i$ and $j,$ and $\bs{a}_i = \norm{\bs{z}_{it}}$ denotes the \textit{activity} of node $i$. The magnitude of the term is proportional to the activity $\bs{a}_i$ of exporting node $i$. 
When $\alpha = 1$, the similarity measure simplifies to $\text{L}_{ijt} = \bs{z}_{it}^\T \bs{z}_{jt}/\norm{\bs{z}_{jt}}$, indicating a global influence of exporting nodes in both networks. Conversely, $\alpha = 0$ reflects a global influence of importing nodes. The parameter $\alpha$ allows $\text{L}_{ijt}$ to smoothly transition between the two extremes, revealing latent characteristics of node influence—whether exporting or importing plays a more significant role in determining edge presence and weight. As a result, the proposed form of $\text{L}_{ijt}$ effectively captures the inherent asymmetry in the trading network. 

One additional feature of \eqref{eq: latent term} is that our inference is invariant to the transpose of the data $\bs{\Delta}_t$ and $\mathbf{Y}_t$. This is in contrast to a more conventional choice of $\text{L}_{ijt} = \bs{z}_{it}^\T \bs{z}_{jt}/\norm{\bs{z}_{jt}}$ corresponds to $\alpha=1$ in \eqref{eq: latent term} \citep{hoff2002}. For this choice, the inference would depend on whether $y_{ijt}$ and $\delta_{ijt}$ encode edge interaction from node $i$ to $j$ (indicating hidden characteristics of exporting countries dominate globally) or in the opposite direction (hidden characteristics of importing countries dominate globally). The proposed formulation \eqref{eq: latent term} eliminates this dependence.

As another significant contribution, Hurdle-Net offers a joint modeling framework by establishing a functional relationship between the two network time series. Specifically, for a monotonically increasing function $g : \bbR \mapsto (0,1)$, it assumes $\bbE(\delta_{ijt})=\bbP(\delta_{ijt} =1) = g ( \bs{x}_{ijt}^\T \, \bs{\beta}_\text{P} + \text{L}_{ijt} )$ and $\bbE(y_{ijt} \con \delta_{ijt} =1) = \bs{x}_{ijt}^\T \, \bs{\beta}_\text{C} + \text{L}_{ijt}$. Given the common latent contribution $\text{L}_{ijt}$, the function $g$ acts as a general link function for modeling the edge occurrence probability.
We set $g(\cdot)$ as a generalized logistic function,
\begin{equation}\label{eq: g}
    g(x) = {\Big( 1 + \exp{(a-bx)} \Big)}^{-1/\gamma}, \quad \mbox{for } a \in \bbR, \mbox{ and } b,\gamma>0 .
\end{equation}
The shared dynamic latent network process, combined with the generalized link function, facilitates the sharing of information between the two time series. This enables Hurdle-Net to effectively and jointly model zero-inflated network time series. Next, we discuss the use of the dynamic shrinkage process \citep{kowal2019} to capture the time-dependence of the hidden characteristics of each node. 


\subsection{Dynamic Shrinkage Process Prior On Latent Variables}\label{Dynamic Shrinkage Process As Prior On Latent Variables}

We first note that Hurdle-Net is not identifiable with respect to $\mathbf{Z}_t$. This is because the latent term $\text{L}_{ijt}$ depends on $\bs{z}_{it}^\T \bs{z}_{jt}$ and $\bs{z}_{it}^\T \bs{z}_{jt} = (\mathbf{P} \bs{z}_{it})^\T \, \mathbf{P} \bs{z}_{jt}$ for any orthogonal matrix $\mathbf{P}$, introducing sign and rotation invariance in latent positions. For identifiability, the common strategy in the literature assumes $\mathbf{Z}_t$ at each time point to be lower triangular with strictly positive diagonals \citep{West2003,Geweke2015}. Instead, at each time $t$ we assume $z_{1t1}$ to be positive, $z_{itk} = \epsilon$ for $k>i$ and a positive $\epsilon$, and leave other $z_{itk}$'s for $i \geq \max (2,k)$ unrestricted; that is
\begin{equation*}
    \mathbf{Z}_t =
    \begin{pmatrix}
        z_{1t1} & \epsilon & \epsilon & \dots  & \epsilon  & \epsilon \\
        z_{2t1} & z_{2t2} & \epsilon & \dots  & \epsilon  & \epsilon \\
        \vdots & \vdots & \vdots & \ddots & \vdots & \vdots \\
        z_{n-1,t,1} & z_{n-1,t,2} & \dots  & \dots  & z_{n-1,t,K-1}  & \epsilon \\
        z_{nt1} & z_{nt2} & \dots  & \dots  & z_{n,t,K-1}  & z_{ntK}
    \end{pmatrix}.
\end{equation*}
It retains all the benefits of the lower triangular assumption, except that at each time point it has only one positive constrained element instead of $K$ many. This improves the mixing of MCMC.

Next, we assume the dynamic shrinkage process (DSP) as the prior on $\{\log z_{1t1}\}_{t=1}^T$ and $\{ z_{itk}\}_{t=1}^T$ for $i \geq \max (2,k)$ \citep{kowal2019}. DSP was motivated by the Bayesian adaptation of trend filtering models \citep{faulkner2018,kim2009,tibshirani2014}. 
For $t = 1, \dots, T$ and $k = 1, \dots, K$, we define
\begin{equation}\label{eq: dhs difference}
\omega_{1t1} = \Delta^d \log z_{1t1} \quad \mbox{and} \quad \omega_{itk} = \Delta^d z_{itk} \quad \forall i \geq \max (2,k),
\end{equation}
where $\Delta^d$ denotes the differencing operator of order $d$ $(\geq 0)$ with $\Delta^0 w_t = w_t$, $\Delta^1 w_t = w_t - w_{t-1}$, $\Delta^2 w_t = \Delta^1 w_t - \Delta^1 w_{t-1}$, and so on. Then DSP hierarchically specifies
\begin{equation}\label{eq: dhs prior}
\begin{split}
    & \omega_{itk} \overset{ind}{\sim} \normal \left( 0, e^{h_{it}} \right),\\
    & h_{i1} = \mu_0 + \mu_i + \eta_{i1},\\
    & h_{it} = \mu_0 + \mu_i + \phi_i \left(h_{i,t-1} - \mu_0 - \mu_i \right) + \eta_{it} \quad \forall t>1,\\
    &  \mu_0, \mu_i, \eta_{it} \,\overset{iid}{\sim}\, Z\left(c, s, 0, 1\right),\\
    & \phi_i \overset{iid}{\sim} \uniform \left( 0, 1 \right).
\end{split}
\end{equation}
Here $Z\left(c, s, \mu_z, \sigma_z\right)$ is the $Z$-distribution as defined in \cite{kowal2019}, which is a general class of distributions encompassing several well-known shrinkage distributions (see Table 1 in \citealt{kowal2019}). For instance, setting $c=s=1/2$, $\mu_z =0$, and $\sigma_z = 1$ results in a horseshoe prior \citep{CARVALHO2010} on $\omega_{itk}$ and corresponds to a dynamic horseshoe process on $\mathbf{Z}_t$, which we use for the rest of the paper. By applying continuous shrinkage to $\omega_{itk}$, a wide range of latent position dynamics can be modeled both flexibly and efficiently. For illustration let $d=1$. Then
\begin{equation*}
\omega_{1t1} = \log z_{1t1} - \log z_{1,t-1,1} \quad \mbox{and} \quad \omega_{itk} = z_{itk} - z_{i,t-1,k} \quad \forall i \geq \max (2,k),
\end{equation*}
with $z_{101} = 1$ and $z_{i0k} = 0$ for $i \geq \max (2,k)$. When differences are 0, the current latent positions are identical to the previous ones. Conversely, a large difference indicates a significant change in latent positions between consecutive time points. When $\omega_{itk}$ is fixed to 0 for all $t \geq 2$, our model degenerates to a static model. The hyperparameters $\mu_0$, $\mu_i$, and $\eta_{it}$ in the prior govern this shrinkage behavior. Specifically, $\mu_0$ and $\mu_i$ regulate the global and node-specific behaviors, while $\eta_{it}$ affects the dynamic shrinkage. An aggressive shrinkage corresponds to $\eta_{it}\approx0$, whereas a large value leads to a considerable change in shrinkage behavior between consecutive time points. DSP defines an AR(1) model for $h_{it}$, allowing the current shrinkage to be influenced by its past values $\{h_{is}\}_{s<t}$. For node $i$, the parameter $\phi_i$ determines this degree of dependence on shrinkage history, with $\phi_i = 0$ indicating no dependence and larger values indicating stronger dependence. We assume independent $\uniform(0,1)$ priors for $\phi_i$ and estimate them from the data.

\subsection{Other Priors And Posterior Inference}\label{subsec: Other Priors And Posterior Inference}

Let $\normal_+$ denote the normal distribution truncated on $(0, \infty)$ and $\gammaprior(s,r)$ denote the gamma distribution with shape $s$ and rate $r$. To complete the hierarchical specification \eqref{eq: cont model}--\eqref{eq: dhs prior}, we impose the following priors on other model parameters,
\begin{equation}\label{eq: other priors}
\begin{split}
    \bs{\beta}_\text{C}, \bs{\beta}_\text{P} \overset{iid}{\sim} \normal( \bs{0}, \sigma_0^2 \mathbf{I}), \quad p(\sigma^2) \propto 1/\sigma^2, \quad \alpha \sim \uniform(0,1),\\
    a \sim \normal(0, \sigma_0^2), \quad b \sim \normal_+ (0, \sigma_0^2), \quad \gamma \sim \gammaprior(s_\gamma, r_\gamma).
\end{split}
\end{equation}
Here, $\mathbf{I}$ denotes the identity matrix. Setting $\gamma=1$ in $g(\cdot)$ (defined in \eqref{eq: g}) reduces it to the standard logistic function. To parameterize the Gamma prior on $\gamma$, we select the smallest integer value for shape parameter $s_\gamma$ so that the mode is at 1. This sets $s_\gamma = 2$ and $r_\gamma = 1$. We choose a large value for $\sigma_0$ to ensure non-informative priors on $\bs{\beta}_\text{C}$, $\bs{\beta}_\text{P}$, $a$, and $b$. For the results presented in Sections~\ref{sec: Simulation Study} and \ref{sec: Bilateral Trade Flows In The Apparel Industry}, we set $\sigma_0 = 10^5$. 
Henceforth, we refer to (\ref{eq: cont model})--(\ref{eq: other priors}) as Hurdle-Net($d$), with $d$ denoting the differencing order in (\ref{eq: dhs difference}).

Let $\bs{\Theta}$ be the collection of all the model parameters. Its posterior distribution, denoted by $p(\bs{\Theta}|\text{data})$, is not analytically available. We use MCMC to sample it from the posterior distribution. In particular, we use the Hamiltonian Monte Carlo algorithm \citep[HMC,][]{neal2011mcmc}. 
HMC is a type of Metropolis-Hastings algorithm utilizes gradient information in the proposal distribution, which can help the Markov chain mix better especially when parameters are highly correlated as in our case. Specifically, an auxiliary momentum variable $\bs{\nu}$ is introduced in HMC. We sample $\bs{\Theta}$ together with the momentum $\bs{\nu}$ from $p(\bs{\Theta},\bs{\nu}|\text{data})=p(\bs{\Theta}|\text{data})p(\bs{\nu})$ where $p(\bs{\nu})\propto \exp(-\bs{\nu}^\T\bs{\Omega}\bs{\nu}/2)$ for some positive definite mass matrix $\bs{\Omega}$. We define the 
potential energy, the kinetic energy, and the Hamiltonian energy respectively as $U(\bs{\Theta})=-\log p(\bs{\Theta}|\text{data})$, $K(\bs{\nu})=-\log p(\bs{\nu})$, and $H(\bs{\Theta},\bs{\nu})=U(\bs{\Theta})+K(\bs{\nu})$. Then the HMC proceeds iteratively as follows: given the current state $(\bs{\Theta},\bs{\nu})$ and the step size $\epsilon$, a new state $(\bs{\Theta}^\star,\bs{\nu}^\star)$ is proposed by applying the following leapfrog scheme several times,
\begin{align*}
    \bs{\nu}\gets \bs{\nu}-\frac{\epsilon}{2}\frac{\partial U(\bs{\Theta})}{\partial \bs{\Theta}},\quad \bs{\Theta}\gets \bs{\Theta}+\epsilon\bs{\Omega}\bs{\nu},\quad \bs{\nu}\gets\bs{\nu}-\frac{\epsilon}{2}\frac{\partial U(\bs{\Theta})}{\partial \bs{\Theta}}.
\end{align*}
The proposed $(\bs{\Theta}^\star,\bs{\nu}^\star)$ is accepted with probability $\min\{1,\exp(-H(\bs{\Theta}^\star,\bs{\nu}^\star)+H(\bs{\Theta},\bs{\nu}))\}$. Since all the parameters are continuous, we implement the HMC using the R
package \texttt{Rstan} \citep{stan2018rstan}. To determine the dimension $K$ of the latent space, we run the MCMC for different values of $K$ and select $K$ that achieves the smallest leave-one-out cross-validation information criterion \citep[LOO-IC,][]{vehtari2015pareto,vehtari2017practical}.

\section{Simulation Study}\label{sec: Simulation Study}

\begin{figure}[!b] 
  \centering
  \includegraphics[width=.9\linewidth]{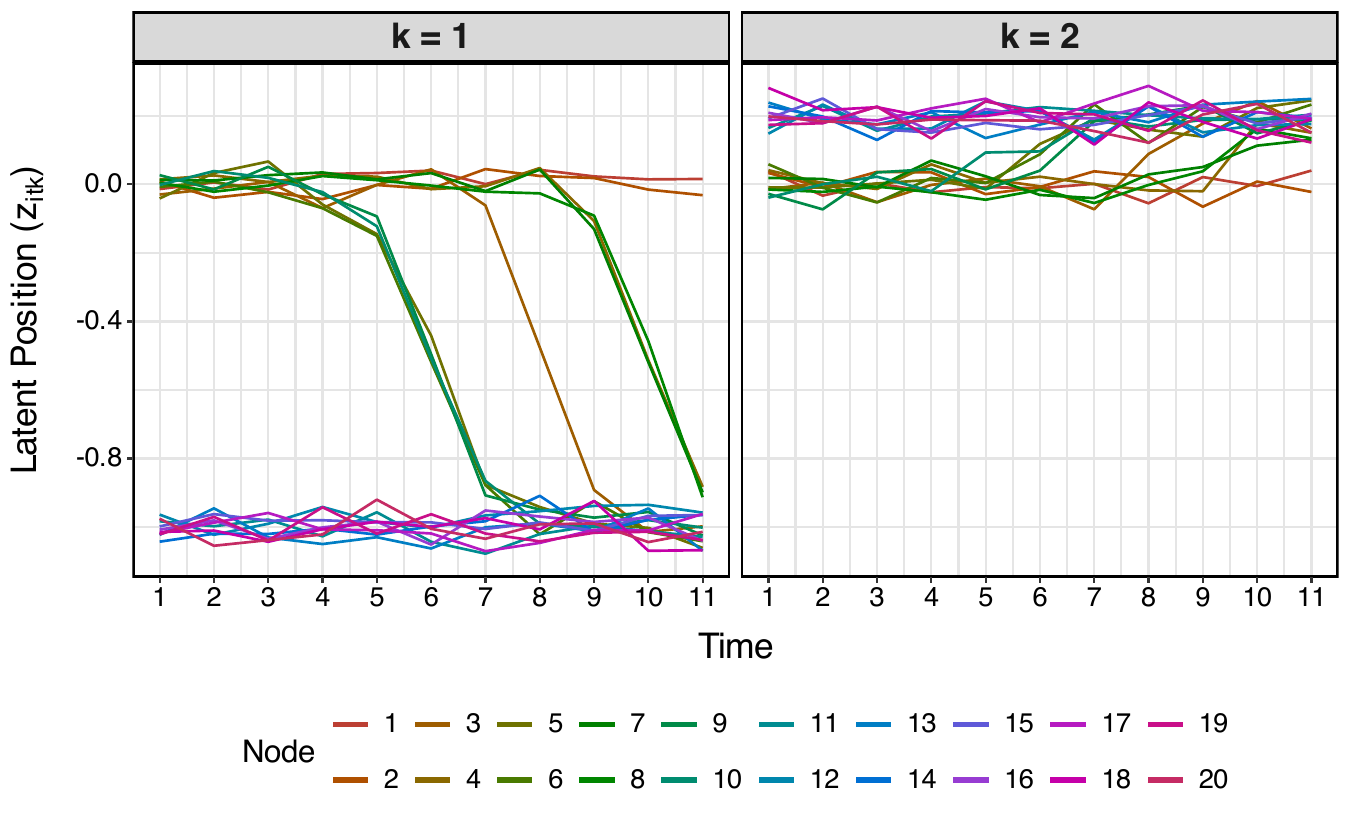}
  \caption{True latent position dynamics in simulations for 11 time points ($T$) for 20 nodes ($n$). Colors indicate different nodes.}
  \label{fig: sim z dynamics}
\end{figure}

In this section, we evaluate the performance of Hurdle-Net through simulation studies. To complement the real-world analysis of the bilateral trade dataset presented in Section~\ref{sec: Bilateral Trade Flows In The Apparel Industry}, we design simulations that mimic its key features. We set the true latent dimension ($K$) to 2. For simulating the latent dynamic network process, we assume two groups of nodes at each $k=1,2$: one with a zero mean and the other with a nonzero mean ($-1$ at $k=1$ and 0.2 at $k=2$). At time $t=1$, the two groups are initially equal in size. Over time, nodes from the zero-mean group gradually shift to the nonzero-mean group, with 80\% of them having transitioned by time $T$.
Figure~\ref{fig: sim z dynamics} illustrates the simulated dynamics for 20 nodes $(n)$ and 11 time points $(T)$.

We vary the number of nodes $n$ as 5, 10, and 20. We fix the number of node-specific covariates ($p_1$) to three and the node-pair-specific covariates ($p_2$) to two. Regression coefficients are set to $\bs{\beta}_\text{C} = (0.0,  0.4, -1,  0.1, -0.7,  0.3, -0.9,  0.6)^\T$ and $\bs{\beta}_\text{P} = (-0.1,  0.3, -0.9,  0.0, -0.6,  0.4, -1.0,  0.7)^\T$. Parameter $\alpha$ in \eqref{eq: latent term} is fixed at 0.9, and node and node-pair specific covariates are randomly simulated from $\normal(0,1)$. The random noise standard deviation ($\sigma$) is set to 1. To mimic real data, we include a single binary node-pair-specific covariate. This covariate is simulated from a $\bernoulli(0.03)$ distribution, matching the observed 3\% occurrence rate of labor provisions over time. For simulating edge occurrence $\delta_{ijt}$, we assume $g$ is the standard normal CDF. For simulation purposes only, we introduce an intercept $\beta_{\text{P}0}$ into the binary model to regulate the proportion of edge occurrences. The analysis model is still \eqref{eq: cont model}--\eqref{eq: probit model} and does not include an intercept. The intercept $\beta_{\text{P}0}$ is set it to 0.5 to match the $70\%$ trade occurrence observed in the real data. For comparison, we consider four different methods:
\begin{description}
    \item[Hurdle-Net(1).] Modeling according to (\ref{eq: cont model})--(\ref{eq: other priors}) with difference order $d=1$. This is the default implementation of our method.
    
    \item[Hurdle-Net(0).] Modeling according to (\ref{eq: cont model})--(\ref{eq: other priors}) with difference order $d=0$. This directly models the latent positions using a continuous shrinkage framework. The model becomes less structured by disabling the shrinkage of current latent positions relative to the previous time point. It solely relies on the log-variance modeling of $h_{it}$ for accommodating the evolution of the latent network structure.
    
    \item[Independent modeling.] This models the edge weight and edge occurrence network time series independently where instead of a common $\text{L}_{ijt}$, we introduce $\text{L}_{\text{C}ijt}$ and $\text{L}_{\text{P}ijt}$ separately for the continuous and binary model. For quantifying the latent contributions, they are defined as in (\ref{eq: latent term}) using model-specific latent positions $\mathbf{Z}_{\text{C}t}$ and $\mathbf{Z}_{\text{P}t}$.
    
    \item[Static modeling.] This assumes that the latent contributions are time-invariant with latent positions $\mathbf{Z} = \left(\bs{z}_{1},\ldots,\bs{z}_{n}\right)^\T$, where $\bs{z}_{i} \in \bbR^{K}$. This is a special instance of Hurdle-Net(1) where $\omega_{itk} = 0$ for all $t \geq 2$. To keep it consistent with the Hurdle-Net, we model it with continuous shrinkage prior, that is, $z_{ik} \overset{ind}{\sim} \normal \left( 0, e^{h_{i}} \right)$ with $h_i = \mu_0 + \mu_i$ and same priors on $\mu_0$ and $\mu_i$.
\end{description}

\begin{figure}[!t] 
  \centering
  \includegraphics[width=\linewidth]{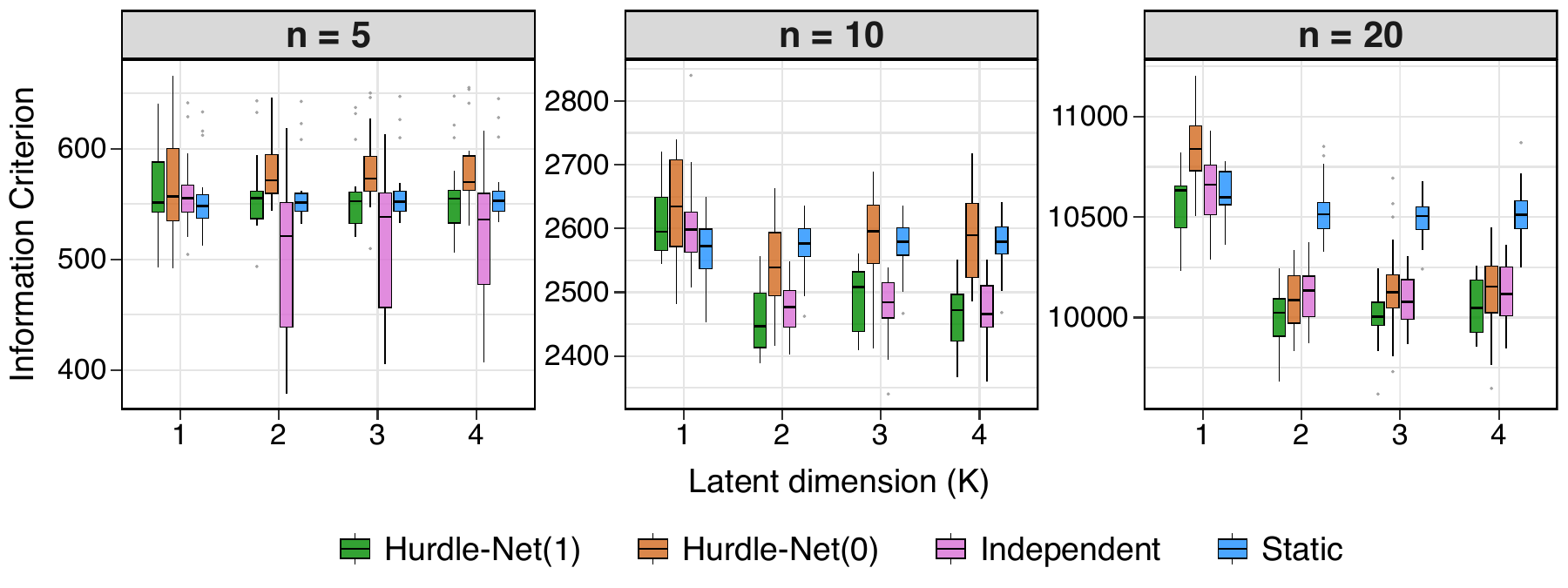}
  \caption{Box-plots of leave-one-out cross-validation information criterion (LOO-IC) values from 20 replications. Panels indicate the different number of nodes $(n)$. Lower values are better.}
  \label{fig: sim ic}
\end{figure}

\begin{figure}[!t] 
  \centering
  \includegraphics[width=.7\linewidth]{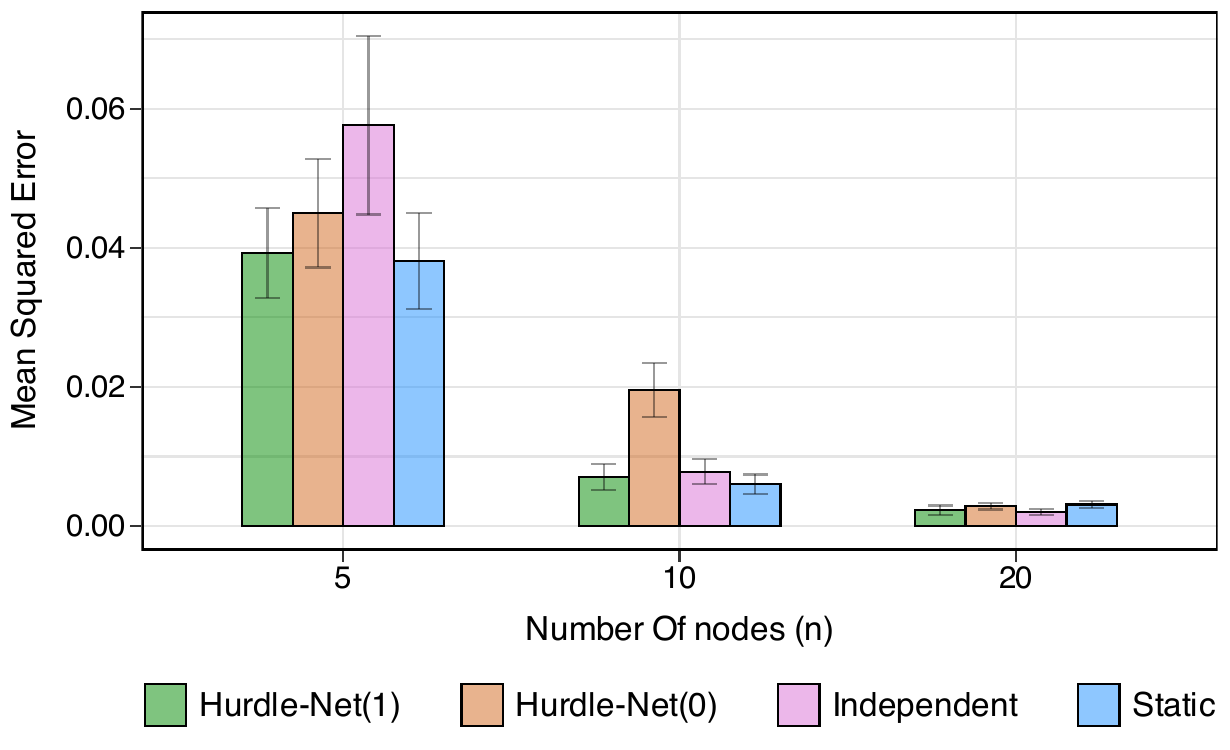}
  \caption{For regression coefficient ($\bs{\beta}_\text{C}$) in the continuous network model (\ref{eq: cont model}), this compares mean squared errors of posterior mean for different numbers of nodes ($n$) in simulations. For this comparison, the methods are fitted with the true value $K=2$. Lower values are better. The error bars indicate $\pm1$ Monte-Carlo standard error across 20 replications.}
  \label{fig: sim betacont mse}
\end{figure}

\begin{figure}[!t] 
  \centering
  \includegraphics[width=.8\linewidth]{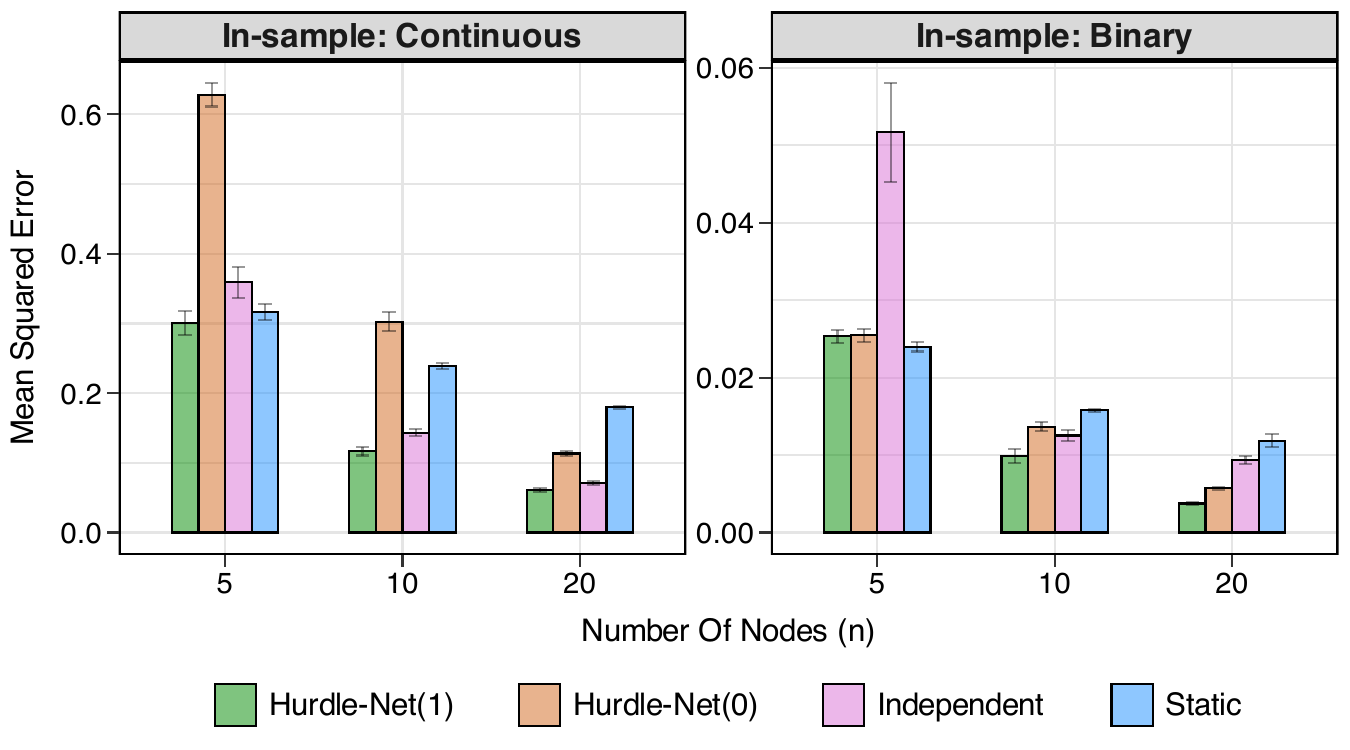}
  \caption{In-sample errors of posterior means of expected edge weights (left panel) and edge occurrence probabilities (right panel) in simulations for different numbers of nodes ($n$). Methods are fitted with the true value $K=2$. Lower values are better. The error bars indicate $\pm1$ Monte-Carlo standard error across 20 replications.}
  \label{fig: sim insample}
\end{figure}

\begin{figure}[!t] 
  \centering
  \includegraphics[width=.8\linewidth]{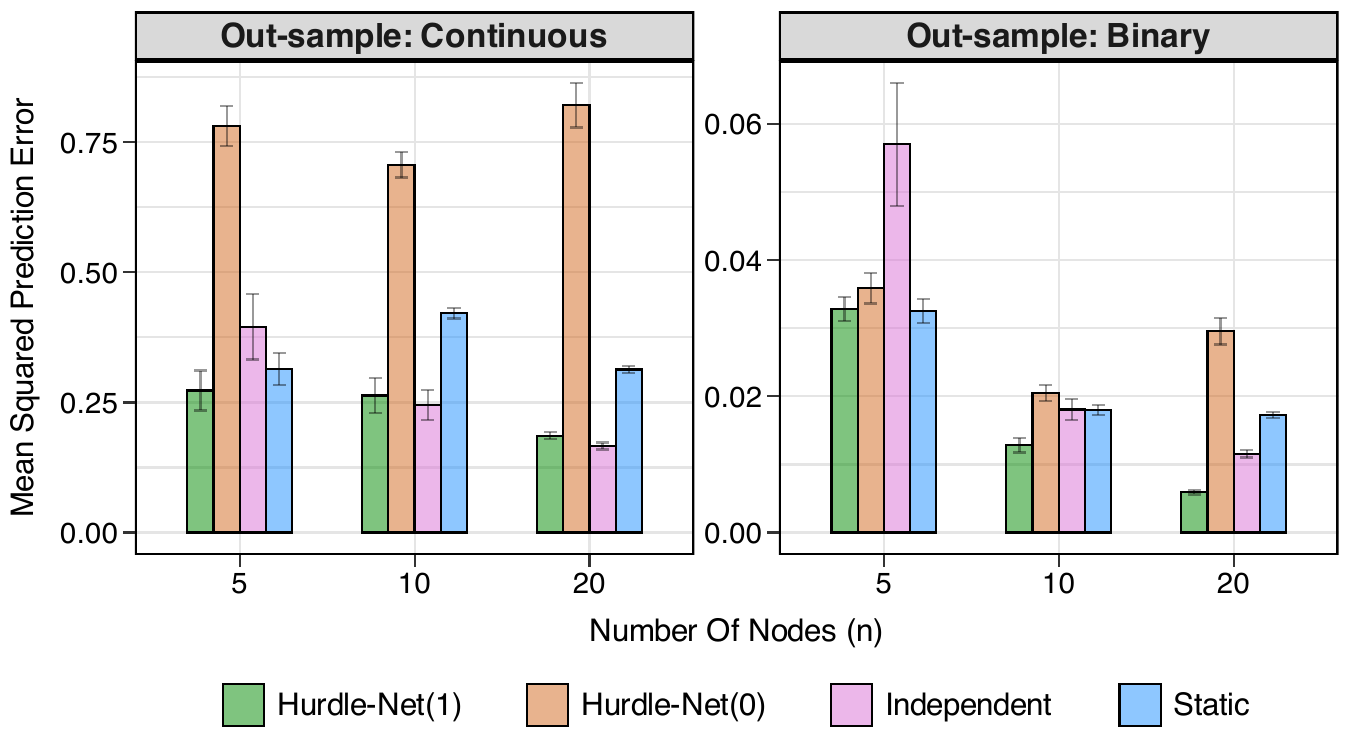}
  \caption{Out-sample prediction errors of predictive median of expected edge weights (left panel) and edge occurrence probabilities (right panel) in simulations for different numbers of nodes ($n$). Methods are fitted with the true value $K=2$. Lower values are better. The error bars indicate $\pm1$ Monte-Carlo standard error across 20 replications.}
  \label{fig: sim outsample}
\end{figure}

We use the simulated data from the first 10 time points for training, and compare their predictive performances at $t=11$. We fit each method with $K = \{1,2,3,4\}$. Figures~\ref{fig: sim ic}--\ref{fig: sim outsample} summarize the performance across the various simulation scenarios, each aggregated over 20 replications. We generate 4,000 MCMC  samples and discard the first 1,000 samples as burn-in.

Figure~\ref{fig: sim ic} presents the leave-one-out cross-validation information criterion (LOO-IC), demonstrating the model selection performance in various scenarios. Overall, the values are lower for Hurdle-Net(1) and the independent model with the minimum occurring at the truth $K=2$. On the other hand, Hurdle-Net(0) and the static model often do not select the true $K$, especially for small $n$. 

Figure~\ref{fig: sim betacont mse} compares the average mean squared error in estimating the regression coefficient (\(\bs{\beta}_\text{C}\)) in the continuous network model. The results show that the estimation error from Hurdle-Net(1) is comparable to that of the best performing method, and this holds true for all values of \(n\).

Figure~\ref{fig: sim insample}--\ref{fig: sim outsample} compare in- and out-sample performances of the methods in estimating and predicting edge weights and edge occurrences. For node pairs in the training data (the first 10 time points), Figure~\ref{fig: sim insample} compares the average mean squared error of posterior means of expected edge weights and edge occurrence probabilities. For node pairs in the test data (the last time point), Figure~\ref{fig: sim outsample} compares the average mean squared prediction error of their predictive medians. The left panel of Figure~\ref{fig: sim insample} shows that Hurdle-Net(1) performs uniformly better than others and becomes similar to the highly parameterized independent method at $n=20$. For estimating edge occurrence probabilities, the right panel of Figure~\ref{fig: sim insample} shows that the independent method performs poorly compared to other methods for small $n$. As $n$ increases, Hurdle-Net(1) outperforms other methods.

For predictive performance, the left panel of Figure~\ref{fig: sim outsample} shows that Hurdle-Net(1) consistently outperforms other methods for small $n$, and matches the highly parameterized independent method as $n$ increases to 20. In predicting edge occurrence probabilities, the right panel shows that the independent method performs poorly for small $n$, while other methods perform similarly. However, as \(n\) increases, Hurdle-Net(1) significantly outperforms the other methods. Overall, Hurdle-Net(1) flexibly adapts to various scenarios, outperforming other methods.

\paragraph{Summary.} The simulation results reveal three main takeaways:
\begin{itemize}
    \item When covariates in a network time series do not capture all dynamic structures in edge weights and occurrences, the latent process is necessary to capture the remaining variability and dependence over time, and a static model is insufficient. Hurdle-Net(1) naturally accounts for this variability and efficiently models the underlying structure.
    
    \item When edge weights and occurrences are influenced by a shared dynamic latent network mechanism, the independent modeling is expectedly less efficient than a joint model. Hurdle-Net offers a comprehensive framework that accommodates this and jointly models the two network time series. This enhances the efficiency of statistical inference.
    
    \item When the latent mechanism has structured dynamics, current latent positions are expected to be related to their immediate past and thus should be informed by them. Hurdle-Net(1), by virtue of the DSP prior, naturally adapts to this. The superior performance of Hurdle-Net(1) over Hurdle-Net(0) validates this approach.
\end{itemize}

\section{Bilateral Trade Flows In The Apparel Industry}\label{sec: Bilateral Trade Flows In The Apparel Industry}

We revisit the bilateral trade flow data from the apparel industry introduced in Section~\ref{sec: Motivating Dataset} and apply Hurle-Net(1), Hurle-Net(0), independent, and static models from Section~\ref{sec: Simulation Study} to analyze them. The data recorded trade occurrences and trade volumes between each pair of $n=29$ countries from 1994 to 2013. In this network, countries are nodes, and the presence of an edge between two countries indicates a trade occurrence. Thus, $\delta_{ijt}$ represents the trade occurrence from country $i$ to $j$ in year $t$, and we define $y_{ijt}$ as the log-transformed nonzero trade volume between them when a trade occurs.

The data include country-specific covariates ($\bs{w}_{it}$): gross domestic product (GDP), population, and area. It also records the distance between capitals, regional trade agreement, and labor provision, which are specific to country pairs ($\bs{w}_{i \bullet j,t}$). Corresponding to the trade observed from country $i$ to $j$ in year $t$, 9 covariates ($\bs{w}_{ijt}$) are thus observed in total, consisting of 3 covariates for the exporter, 3 for the importer, and 3 for the country pair. Regional trade agreement and labor provision are binary variables with 1 indicating the presence of agreements. The remaining covariates are continuous.

\begin{figure}[!t]
     \centering
     \includegraphics[width=.85\linewidth]{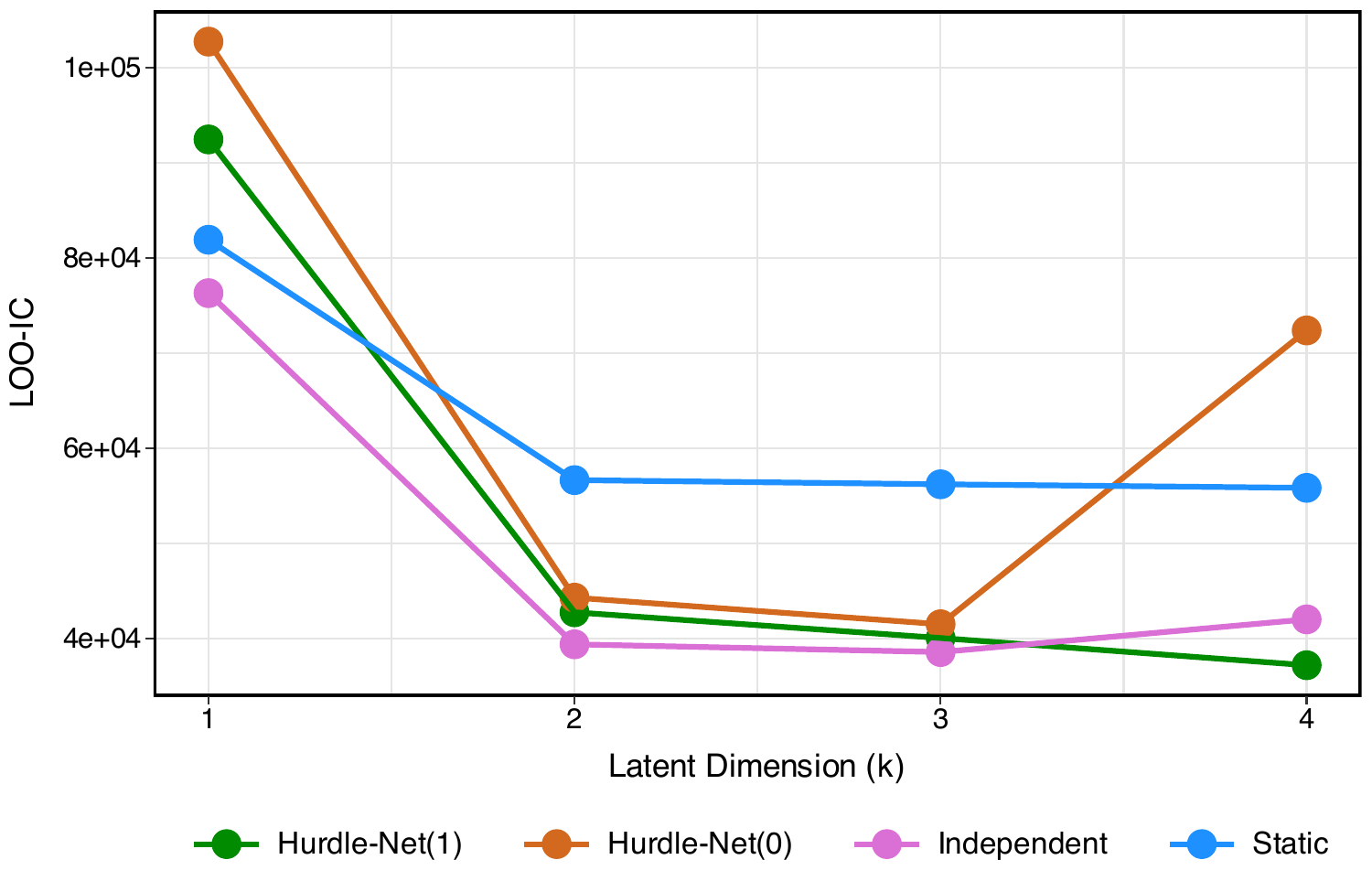}
   \caption{The leave-one-out cross-validation information criterion (LOO-IC) in bilateral trade flow analysis using different methods for a varied range of latent dimensions ($K$). Lower values are better.}\label{fig: trade looic}
\end{figure}

\begin{figure}[!t]
     \centering
     \includegraphics[width=.97\linewidth]{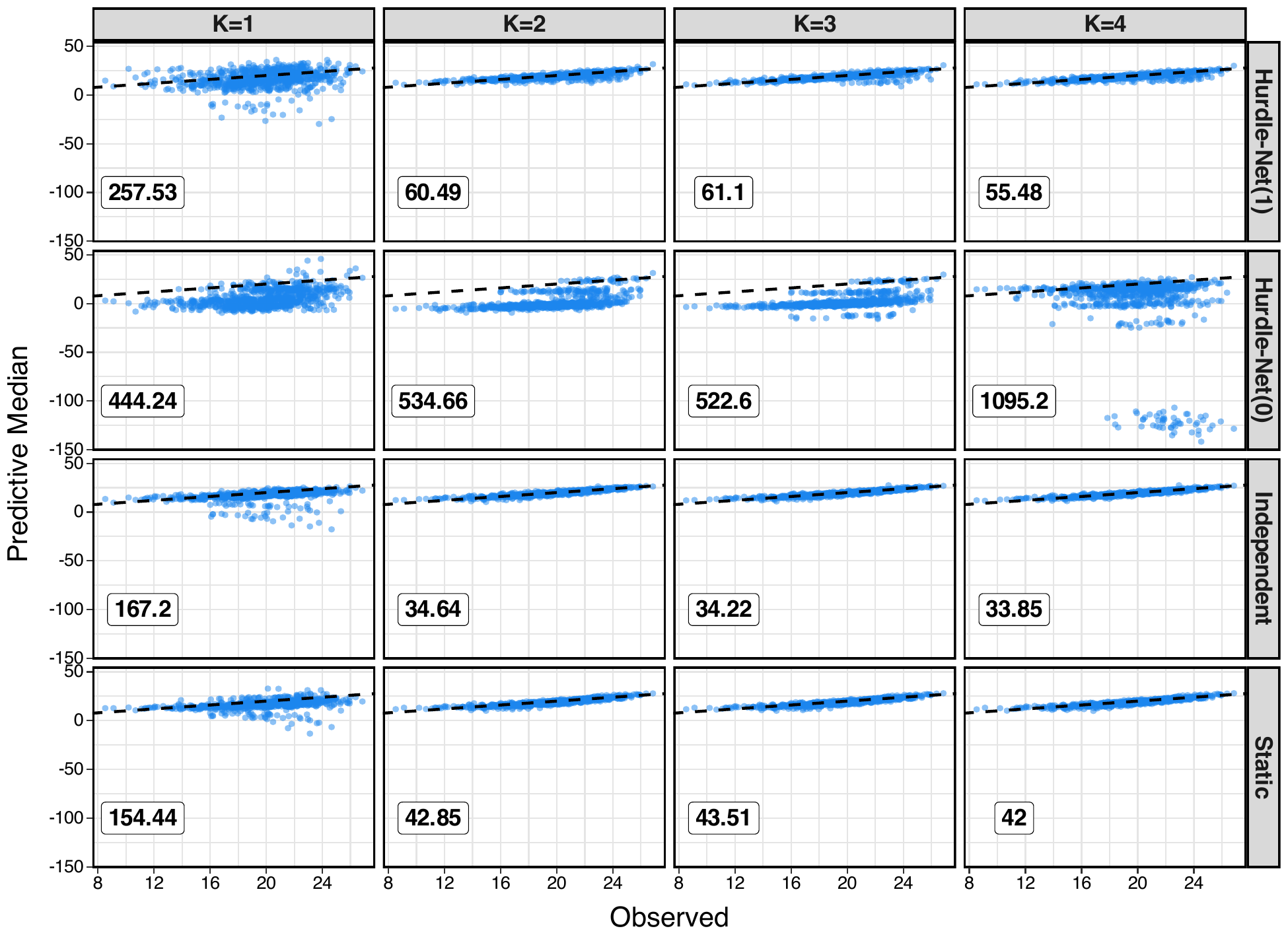}
   \caption{For a varied range of latent dimensions ($K$; column), this displays log nonzero trade volume predictions (predictive median) in 2013 from different models (row) for country pairs for which trades occurred. The horizontal and vertical axes display the observed and predictive medians of log-transformed nonzero trade volumes. The mean squared prediction error is shown in the bottom-left corner of each panel. Lower values indicate better performance. The dashed black line represents the $y=x$ line.}\label{fig: trade volume}
\end{figure}
We use the data from 1994 to 2012 to train the models and compare their predictions in 2013. We fit each method with $K = \{1,2,3,4\}$. Since the range of continuous covariates varies widely, we standardize them. Let $m_l$ and $s_l$ denote the sample mean and standard deviation of the $l$-th covariate in the training data combined across country pairs and time points. We set $x_{ijtl} = (w_{ijtl} - m_l) / s_l$ if the $l$-th covariate is continuous, and leave it unchanged (i.e., $x_{ijtl} = w_{ijtl}$) for the binary covariate. We generate 6,000 MCMC  samples and discard the first 4,000 samples as burn-in.

\paragraph{Model selection.} Figure~\ref{fig: trade looic} depicts the LOO-IC for the four methods across different latent dimensions ($K$). For all methods, the most significant decrease in LOO-IC occurs at $K=2$. While further increasing $K$ has no effect for the static model, the LOO-IC for independent and Hurdle-Net(0) increases at $K=4$, with Hurdle-Net(1) achieving the best model selection (lowest value).

\begin{figure}[!t]
     \centering
     \includegraphics[width=.95\linewidth]{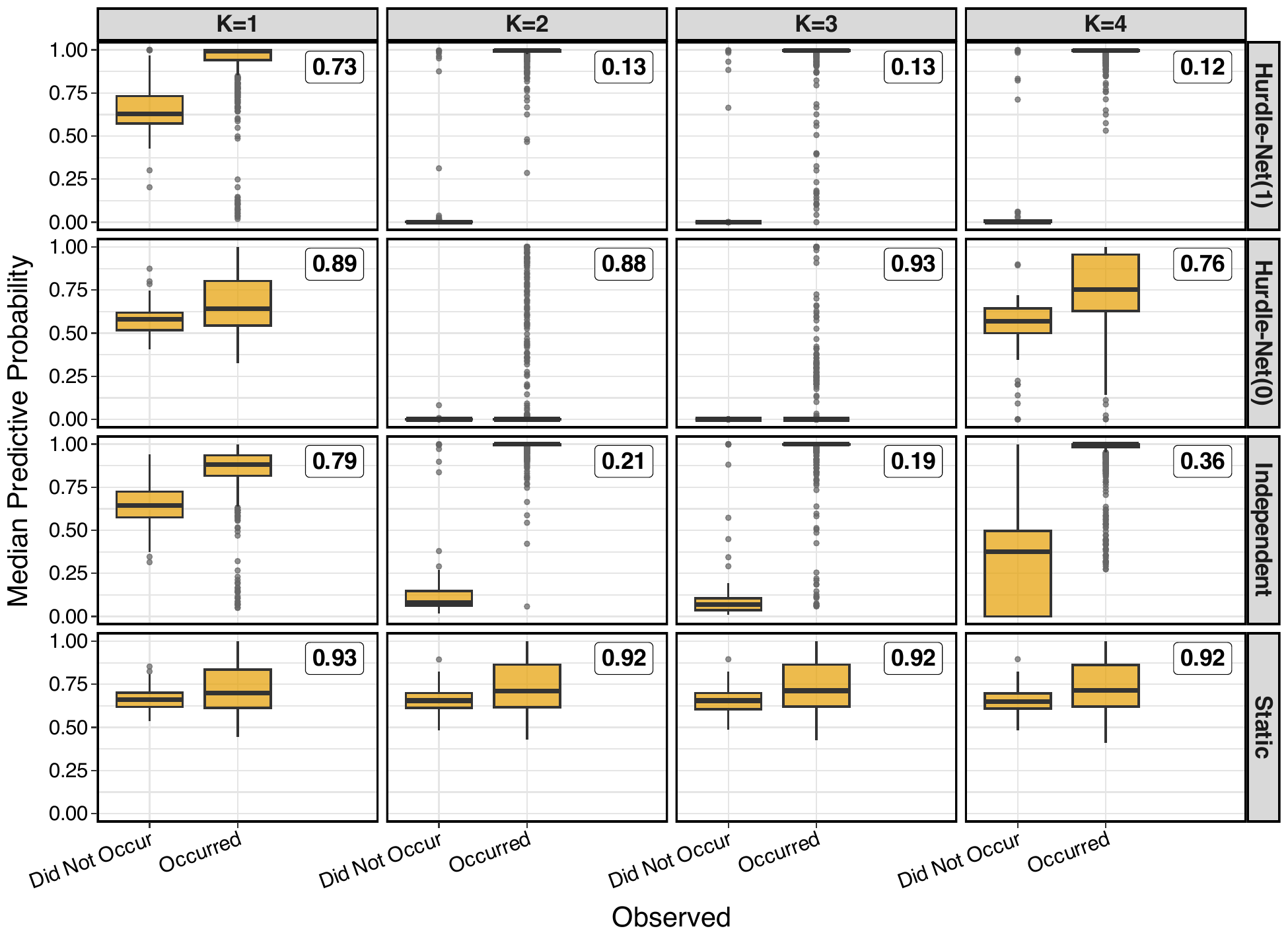}
   \caption{For a varied range of latent dimensions ($K$; column), this displays trade occurrence predictions (predictive median probability) in 2013 from different models (row) for all country pairs. The horizontal axis displays the observed trade occurrence status, and the vertical axis shows median predictive probabilities from modeling. The average absolute prediction error \eqref{eq: occurrence error} is shown in the top-right corner of each panel. Lower values indicate better performance.}\label{fig: trade occurrence}
\end{figure}

\paragraph{Trade volume and trade occurrence predictions in 2013.} Figures~\ref{fig: trade volume} and \ref{fig: trade occurrence} compare the log trade volume and trade occurrence predictions (predictive median of expected log nonzero trade volume and trade occurrence probability) in 2013 from different models across various latent dimensions ($K$). Log trade volume predictions from Hurdle-Net(1), independent, and static models show significant improvement at $K=2$. While further increases in $K$ provide substantial benefits for Hurdle-Net(1), peaking at $K=4$, the improvements for independent and static models are negligible. The performance of Hurdle-Net(0) is poor for all $K$ because the predictive distribution of latent positions (marginalized with respect to their variance) is symmetric about 0.

Let $\delta_{ij}$ and $\hat{\delta}_{ij}$ denote the observed trade occurrences and their predictive median probabilities in 2013. To compare them, we compute the error measure
\begin{equation}\label{eq: occurrence error}
    \frac{1}{N_1} \mathop{\sum \sum}_{i \neq j, \, \delta_{ij} = 1} \abs{\hat{\delta}_{ij} - 1} + \frac{1}{N - N_1} \mathop{\sum \sum}_{i \neq j, \, \delta_{ij} = 0} \abs{\hat{\delta}_{ij}},
\end{equation}
where $N = n(n-1)$ is the total number of country pairs, and $N_1$ is the number of country pairs for which trade occurred. The measure ranges from 0 (perfect prediction) to 2 (worst prediction) and quantifies the average absolute errors in predicting the presence and absence of trades. This is presented in Figure~\ref{fig: trade occurrence}. Trade occurrence predictions from Hurdle-Net(1) and independent models show the most improvement at $K=2$. A further increase in $K$ improves Hurdle-Net(1)'s predictions for $K \geq 3$, significantly outperforming the independent model. The static model's predictions remain unchanged across different $K$ values, while Hurdle-Net(0) performs poorly, being overly conservative in predicting the presence of trades.

\begin{figure}[!t]
     \centering
     \begin{subfigure}[b]{\linewidth}
         \centering
         \includegraphics[width=\linewidth]{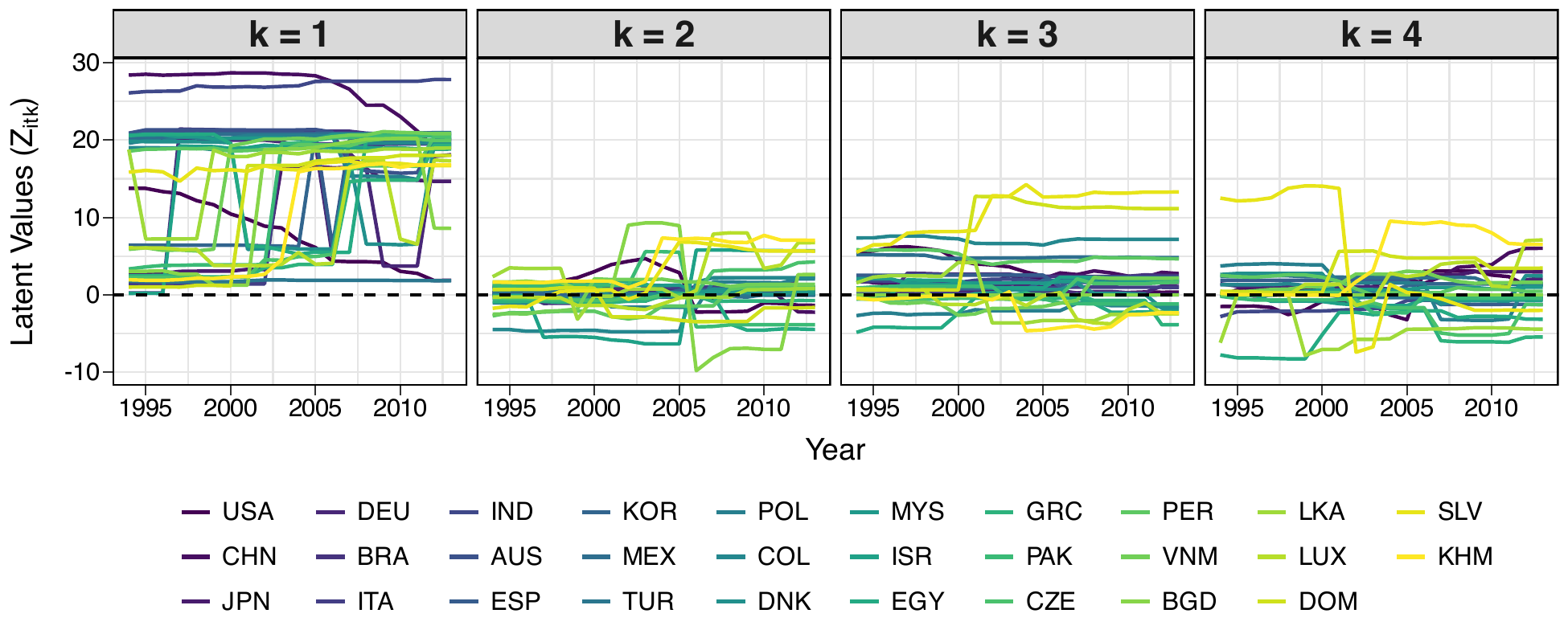}
         \caption{Dynamics of latent positions}\label{subfig: trade latent position dynamics}
     \end{subfigure}
     \begin{subfigure}[b]{\linewidth}
         \centering
         \includegraphics[width=\linewidth]{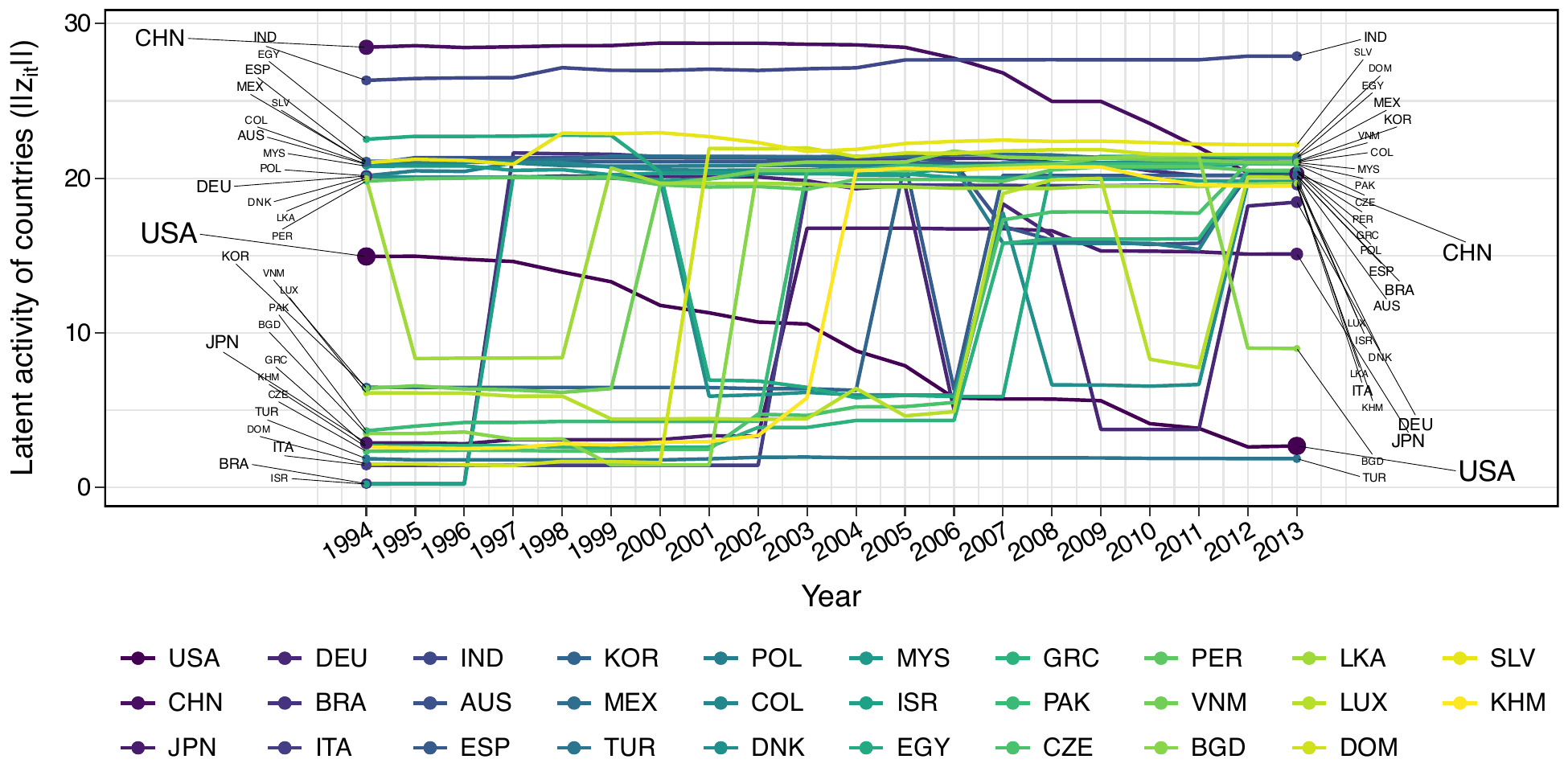}
         \caption{Dynamics of the Euclidean norm (activity) of latent positions}\label{subfig: trade latent norm dynamics}
     \end{subfigure}
   \caption{Estimated latent dynamics from Hurdle-Net(1) with $K=4$. The top panel (a) plots estimated latent positions for each country over time. The bottom panel (b) plots their Euclidean norm over time, indicating latent activity of countries. The values plotted for 2013 are predictions. Countries are color-coded from darker to lighter shades according to decreasing GDP observed in 2013.}\label{fig: trade latent dynamics}
\end{figure}

\paragraph{Summary.} The findings from Figures~\ref{fig: trade looic}--\ref{fig: trade occurrence} can be summarized as follows.
\begin{itemize}
    \item The dynamic models outperform the static model in both model selection performance and trade prediction. This indicates strong evidence for a temporal dependence in the latent network mechanism and underscores the importance of its dynamic modeling to more accurately capture the trade mechanism.
    
    \item Hurdle-Net(1) consistently outperforms Hurdle-Net(0) in all aspects. Its performance significantly improves by anchoring the current latent position to its value at the previous time point, indicating the potential presence of structured latent dynamics. Hurdle-Net(1), leveraging the DSP prior, naturally adapts to this and proves to be more suitable and efficient for modeling the trade mechanism.

    \item Hurdle-Net(1) performs comparably to the independent model in terms of nonzero trade volume prediction, but it excels in model selection and predicting trade occurrence, improving the latter error measure by about $66\%$. This is noteworthy since Hurdle-Net(1) achieves this with significantly fewer parameters. This validates the assumption of a common dynamic latent network structure underlying both trade occurrence and nonzero trade volume network time series. By jointly modeling these two network time series, Hurdle-Net(1) achieves both parsimony and efficiency, thereby enhancing the effectiveness of statistical inference. Because of these advantages, we focus on interpreting the results from Hurdle-Net(1) for the rest of this section.
\end{itemize}

\begin{figure}[!t]
     \centering
     \includegraphics[width=.9\linewidth]{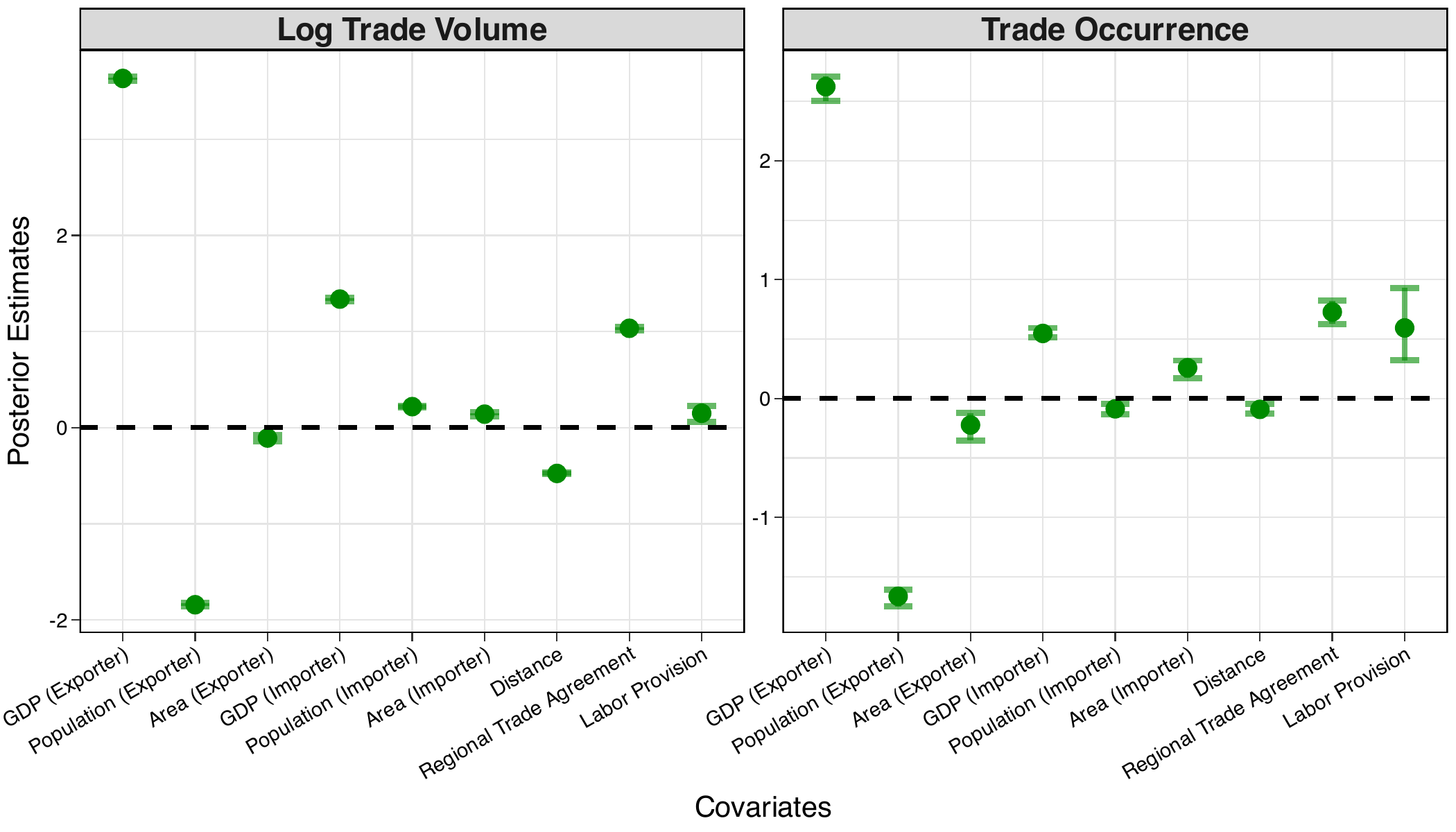}
   \caption{Regression coefficient estimates in continuous ($\bs{\beta}_\text{C}$; left panel) and binary ($\bs{\beta}_\text{P}$; right panel) network models for $K=4$. The uncertainties around point estimates are indicated by $95\%$ credible intervals. The horizontal dashed black line is $y=0$ and indicates no influence.}\label{fig: trade reg coeff}
\end{figure}

\paragraph{Estimated latent dynamics and regression coefficients.} Figure~\ref{fig: trade latent dynamics} presents the latent dynamics estimated using Hurdle-Net(1) for $K=4$. Figure~\ref{subfig: trade latent position dynamics} illustrates the estimated latent positions of countries over time. The primary contribution to the latent dynamics comes from the first dimension, with its values approximately ranging between 0 and 20. 
Figure~\ref{subfig: trade latent norm dynamics} summarizes variations across latent dimensions by plotting the dynamics of the Euclidean norm (activity) of latent positions for each country. In 1994, two broad clusters of activity (below and above 15) are initially observed, each containing approximately equal numbers of countries. Over time, more countries transition to the higher activity cluster, leaving the United States (USA), Bangladesh (BGD), and Turkey (TUR) in the lower activity cluster. In (\ref{eq: cont model})--(\ref{eq: probit model}), covariates and latent variables together aim to explain the bilateral trade flow. The patterns in the estimated latent dynamics indicate that, in 1994, covariates effectively explain the trade patterns, requiring significant latent contributions for half of the countries. As time progresses, more countries require contributions from latent variables, and they vary over time. This underscores the importance of incorporating a dynamic latent network mechanism, as implemented in Hurdle-Net(1).

Given covariates and latent variables together explain the bilateral trade flow, Hurdle-Net(1) by virtue of the dynamic latent modeling naturally adjusts the importance of covariates. This leads to more accurate estimates of the regression coefficients ($\bs{\beta}_\text{C}$) and ($\bs{\beta}_\text{P}$) for explaining nonzero trade volumes and trade occurrences. Figure~\ref{fig: trade reg coeff} presents these estimates from Hurdle-Net(1) for $K=4$. It is evident that the GDP of exporters has the most significant positive effect on both nonzero trade volumes and occurrences. The GDP of importers and a regional trade agreement also positively impact both. Moreover, the statistical significance of labor provision in determining trade occurrence is consistent with findings in the Economics literature \citep{leclercq2020labor}. Other covariates show near-null effects, except for the negative effects of the population of exporters (for both trade occurrence and nonzero trade volumes) and the distance between countries (for nonzero trade volumes).

\section{Discussion}\label{sec: Discussion}

With technological advancements, network time series data have become increasingly prevalent, recording interactions among a set of individuals over time. Motivated by bilateral trade flow data as a representative example, this article introduces a novel Hurdle-Net as a structured modeling framework. About $30\%$ of country pairs in the data record no trade occurrences, making it an example of zero-inflated directed network time series. The data can be represented as a paired continuous and binary directed network time series, which together capture both trade occurrences and nonzero trade volumes. Also, there is an inherent dependence between the two networks since the same countries are involved. 

The proposed Hurdle-Net utilizes a latent dynamic shrinkage process for improved edge occurrence prediction. The method jointly and dynamically models binary and continuous directed network time series, and includes several novel components. Assuming node-specific latent variables, we introduce a latent term similar to that in \cite{hoff2002} to quantify the latent contribution. It retains the benefits of \cite{hoff2002} while ensuring that the model remains invariant to the transpose/orientation of the data. By using a common latent mechanism and a generalized logistic function as a general link function, Hurdle-Net creates a parsimonious functional relationship between the two network time series. This allows us to model the two time series jointly. Unlike independent modeling approaches similar to prior research in \cite{ward2013}, Hurdle-Net implicitly assumes a common latent mechanism governing both networks and models them jointly. The dynamic shrinkage process (DSP) prior on latent positions captures underlying structures in the latent dynamic evolution. By using a global-local continuous shrinkage prior, DSP comprehensively models dynamic changes and flexibly accommodates a wide range of latent dynamic processes.

Our approach uncovers valuable findings through the estimates of interpretable model parameters. The estimated latent dynamics reveal the underlying network structure and allow us to analyze variations in activity and the latent contributions of different countries over time. This allows Hurdle-Net to capture temporal dependencies, the relationship between edge occurrence and weights, as well as network dependencies among nodes. By inherently accounting for the underlying latent structure, Hurdle-Net delivers  accurate estimates of the importance of various covariates. Compared to methods that do not incorporate trade-volume data (independent model) or time-dependence (static model), this leads to enhanced inference on edge occurrence while maintaining similar performance in edge weight inference.

In many applications, the underlying mechanisms of time series data involving the same set of individuals are often unclear. The use of a dynamic latent network structure is valuable in this context, offering broad applicability across various scientific fields, far beyond just bilateral trade flows. Examples include functional connectivity networks among brain regions \citep{Tomasi2011}, interactions within social networks \citep{Wilson2009}, email communication networks \citep{diesner2005exploration}, citation networks among research articles or authors \citep{Yan2012}, and networks of co-purchased products \citep{Kafkas2021}. In these contexts, the latent dynamic modeling of network structures that we propose here will be useful. By efficiently accommodating latent structures, it can enhance the estimation and prediction of edge occurrences and continuous values between node pairs. 



\section{Funding}\label{sec: Funding}
Ni's research was partially supported by NSF DMS-2112943.



\section{Software}\label{sec: Software}
The codes for implementing the framework and reproducing simulations can be found in the GitHub repository ``hurdlenet'' at \url{https://github.com/sandy-pramanik/hurdlenet}. For future updates, please refer to this repository.

\bibliographystyle{apalike}
\bibliography{references}

\end{document}